\documentclass[prl,twocolumn,showpacs]{revtex4}

\usepackage{graphicx}
\newcommand{\etal}{\textit{et al.}}
\newcommand{\beq}{\begin{equation}}
\newcommand{\eeq}{\end{equation}}
\newcommand{\tc}{\theta_\mathrm{c}}
\renewcommand{\t}{\theta}
\newcommand{\sLS}{\sigma_\mathrm{LS}}
\newcommand{\sGB}{\sigma_\mathrm{GB}}
\newcommand{\Peq}{P_\mathrm{eq}}
\newcommand{\Tm}{T_\mathrm{m}}
\newcommand{\PS}{P_\mathrm{S}}
\newcommand{\PL}{P_\mathrm{L}}

\newcommand{\rS}{\rho_\mathrm{S}}
\newcommand{\rL}{\rho_\mathrm{L}}
\newcommand{\lc}{l_\mathrm{c}}

\begin{document}

\title{Wetting properties of grain boundaries in solid helium 4}
\author{Satoshi Sasaki, Fr\'ed\'eric Caupin, and S\'ebastien Balibar}
\affiliation{Laboratoire de Physique Statistique de l'Ecole Normale
Sup\'erieure\\
associ\'e au CNRS et aux Universit\'es Paris 6 et Paris 7, 24 rue
Lhomond 75231 Paris Cedex 05, France}
\begin{abstract}
We have observed boundaries between hcp $^4$He crystal grains in
equilibrium with liquid $^4$He.  We have found that, when emerging at the
liquid-solid interface, a grain boundary makes a groove whose dihedral
angle $2\theta$ is non-zero.  This measurement shows that grain
boundaries are not completely wet by the liquid phase, in agreement 
with recent Monte Carlo simulations.  Depending on the value of 
$\theta$, the contact line of a grain boundary with a
solid wall may be wet by the liquid.  In this case, the line is a thin
channel with a curved triangular cross section, whose measured width
agrees with predictions from a simple model.  We discuss these
measurements in the context of grain boundary premelting and for a
future understanding of the possible supersolidity of solid $^4$He.
\end{abstract}
\pacs{67.80.-s, 61.72.Mm, 61.30.Hn, 68.35.-p}
\maketitle

Kim and Chan~\cite{Kim1,Kim2} have discovered anomalies in the
behavior of a torsional oscillator (TO) containing solid $^4$He, and
interpreted them as evidence for ``supersolidity'', i.e. superfluidity
in a solid state.  Using the same method, Rittner and Reppy have
observed similar anomalies but shown that they disappeared after
annealing~\cite{Rittner06} and could be enhanced in quickly grown
solid samples~\cite{Rittner07}.  This suggests that, if supersolidity
exists, it is not an intrinsic property of $^4$He crystals but rather
a consequence of disorder in samples which can be polycrystalline or
even glassy.  In a recent experiment, we observed dc-mass flow through
solid $^4$He at the liquid-solid (LS) equilibrium, but only in the
presence of grain boundaries (GBs)~\cite{Sasaki06}.  The whole issue
of supersolidity in $^4$He being still controversial, it is important
to study the physics of GBs in solid $^4$He.  Monte Carlo
simulations~\cite{Pollet} predict that GBs have a microscopic
thickness, about 3 atomic layers: they should not be completely wet by
the liquid phase at the LS equilibrium; it is also predicted that GBs
are superfluid below around $0.5\,\mathrm{K}$, except for special
orientations.  Other numerical studies are in
progress~\cite{Ceperley07,Reatto07}.  In $^4$He, GBs thus appear as
interesting quasi-2D quantum systems to be studied experimentally.

When emerging at the liquid-solid interface, a grain boundary makes a
groove with a dihedral angle $2\theta$; we have found that $2\theta$
is non-zero.  This proves that grain boundaries are not completely
wet by the liquid phase at the LS equilibrium.  From the
angle $\theta$, we deduce a grain boundary energy $\sGB$ which is
smaller than twice the liquid-solid interfacial energy $\sLS$.  We
also show that the contact line of GBs on glass windows may or may not
be wet by the liquid, depending on $\theta$ and in agreement with
a simple model.  The premelting of GBs near the solid-liquid
transition is an important issue in materials
science~\cite{Besold94,Dash95} and direct experimental evidence of GB
premelting is scarce in pure systems; it might occur very close to the
melting point in aluminum~\cite{Balluffi} and in colloidal
crystals~\cite{Alsayed05}.  Our study shows that there is no
premelting in $^4$He, except close to a wall.

GBs in $^4$He had already been observed by Franck~\etal, but at high
temperature and above $50\,\mathrm{MPa}$~\cite{Franck83,Franck85}.
Franck \etal~studied films about $50\,\mathrm{\mu m}$ thick with a
Schlieren method.  For the fcc phase, a polygonal, foam-like structure
was observed; upon heating, the vertices of the foam widened into a
curved triangular shape, before melting occurred at the GBs, the
grains of the crystal becoming separated.  Measurement of dihedral
angles was difficult and gave $0\textsuperscript{o}
\leq \theta \leq 30\textsuperscript{o}$.  Because some invasion of the
GBs by the liquid was observed together with grain separation upon
heating, this experiment is cited by Dash \etal~as an evidence for
near complete wetting~\cite{Dash95} or even premelting~\cite{Dash05}
of the GBs.  However, Franck~{\etal} studied foam-like thin films,
with one side in contact with a glass window, and the other with the
liquid.  As explained below, our results question Dash's
interpretation of Franck's observations.  Franck \etal~also observed
the hcp phase which exhibited banded structures with no GB melting,
but they could not measure $2\theta$~\cite{Franck85}.

We have made a square cell by closing with two glass windows a
$11\,\mathrm{mm} \times11\,\mathrm{mm}$ hole in a $3\,\mathrm{mm}$
thick copper plate.  This plate is cooled by a dilution refrigerator
with optical access along a horizontal axis.  The cell allows to
visualize the contact of helium crystals between themselves, with
copper walls (on the sides) or glass walls (the two windows and a
glass piece glued at the top).  Images were taken with a CCD camera
outside the cryostat.  For all crystals studied here, the temperature
was between 40 and 100$\,\mathrm{mK}$.

\begin{figure}
\centerline{\includegraphics[width=1\linewidth]{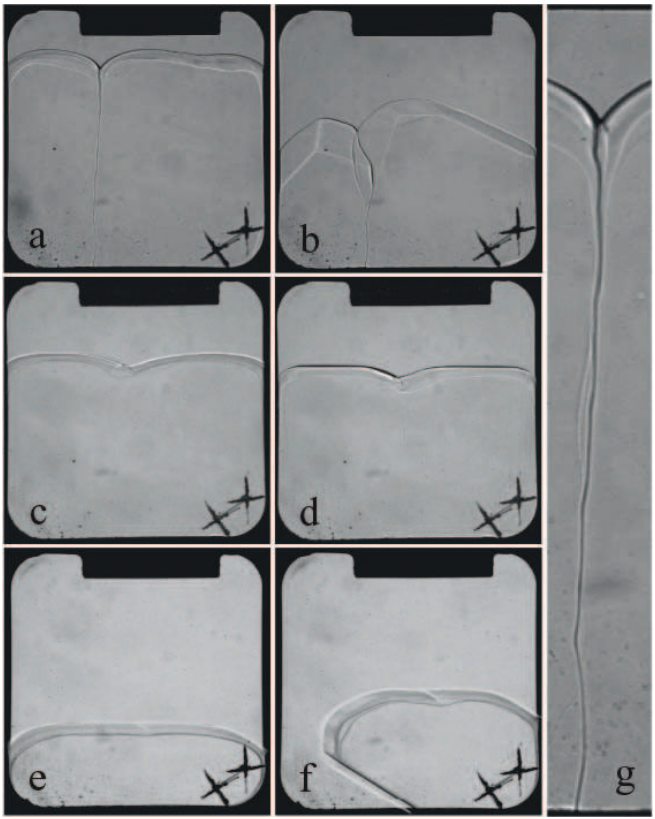}}
\vspace{0.5mm}
\caption{Three pairs of images showing equilibrium shapes (a,c,e)
together with growth shapes which reveal the crystal orientation
(b,d,f).  When the two crystal grains have a large difference in
orientation (a,b), their boundary ends as a deep groove at the
liquid-solid interface.  A zoom of (a) shows that the contact lines of
the GB with the windows are in fact liquid channels (g).  Crystals
with similar orientations can be obtained by direct growth (c,d,e,f).
In this case, the groove is shallow with no liquid channels on the
windows.  Two crosses carved on the windows (lower right corners) help
adjusting the focusing.}
\label{f1}
\end{figure}

When grown slowly by pressurizing superfluid liquid below 1~K, solid
$^4$He is usually a good quality single crystal~\cite{Balibar05}.  To
make polycrystals we could either solidify $^4$He at constant volume
from the normal liquid at high temperature or inject mass at low
temperature at a fast rate (even when crystallizing the whole sample
in 0.1~s, the fill line usually remained open~\cite{SasakiJLTP}).  In order to avoid
lengthy temperature cycles, we chose the latter method.  By slowly
releasing the pressure, we could melt part of it and reveal its
polycrystalline structure.  Further melting and growth lead to two
grains with a single GB as shown on Fig.~\ref{f1}(a).  The temperature
being highly homogeneous, gravity is relevant and the solid occupies
the bottom of the cell.

\begin{figure}
\centerline{\includegraphics[width=1\linewidth]{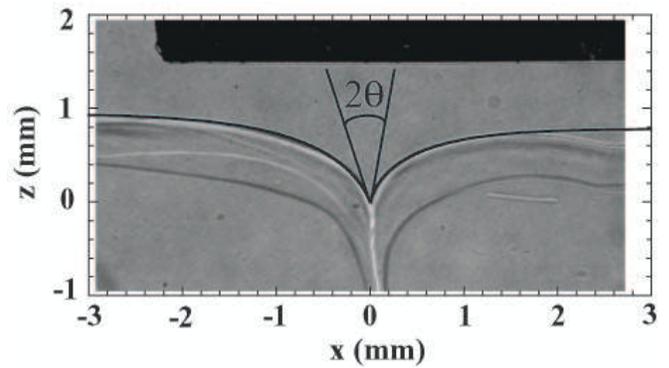}}
\caption{The cusp angle $\theta$ is determined by fitting each crystal profile 
 with the Laplace equation near the cusp \cite{note}.}
\label{f2}
\end{figure}

Figs.~\ref{f1}(a,c,e) show that, when emerging at the LS
interface, a GB makes a groove whose angle $2\theta$ is non-zero. Care was taken to
view the groove along the GB direction: the optical axis was rotated 
until the lowest $\theta$ was found; because the cell is thin, the GBs
are often nearly perpendicular to the windows.  The groove results
from the mechanical equilibrium between the GB surface tension
$\sigma_\mathrm{GB}$ and the LS surface tension $\sigma_\mathrm{LS}$.
It is known that $\sigma_\mathrm{LS}$ varies from 0.16 to 
0.18~mJ/m$^2$ depending on orientation~\cite{Balibar05}. 
If, for simplicity, one neglects this
anisotropy, mechanical equilibrium requires
\beq
\sigma_\mathrm{GB} =
2\,\sigma_\mathrm{LS}\,\cos{\theta} \, ,
\label{e1}
\eeq
where $\theta$ is half the dihedral angle of the groove.  In the case
of Fig.~\ref{f1}(a), the optical axis is slightly misaligned in order
to show the existence of two liquid channels along the windows.  For
three well aligned GBs with various orientations, we zoomed on the
cusp to measure $\theta$.  They correspond to the general case where
the difference in orientation between the two grains is large, as
shown by growth shapes (Fig.~1(b)).  We obtained $\theta\, = \,
11\,\pm\,3$, $16\,\pm\,3$, and $14.5\,\pm\,4 \, \textsuperscript{o}$
by fitting the respective crystal profiles near the cusp with a one dimensional
Laplace equation, assuming that the curvature in the plane
perpendicular to the windows was constant (see
Fig.~\ref{f2})\cite{note}.

In the case of Figs.~\ref{f1}(c,d,e,f), crystals were obtained by
direct growth, not by melting a foam.
Fig.~\ref{f1}(d) shows two crystals with parallel c-facets, but it is
possible that the other facets are not.  The GB has a lower energy as
shown by a larger groove angle.  Figs.~\ref{f1}(e,f) show a very
shallow groove; they might correspond to a single crystal with a
stacking fault.

From Eq.~(\ref{e1}), the GB energy is easily obtained as a function of
$\sLS$.  The depth $\Delta z$ of the groove is related to $\theta$
through
\beq
(\rS-\rL)\,g\,(\Delta z)^2 = 2\,\sLS\,(1-\sin{\theta})
\label{e2}
\eeq
with $\rho_\mathrm{L}$ (resp.  $\rho_\mathrm{S}$) the mass per unit
volume of the liquid (resp.  the solid), and $g$ the gravity.
Eq.~(\ref{e2}) leads to values of $\theta$ which are consistent with
those which are directly measured.  Note that, in materials science, a
similar method is widely used to measure
$\sLS$~\cite{Bolling60,Boyuk07}, where the effect of gravity is
replaced by a thermal gradient.  In the case of $^4$He, $\sLS$ is
known from direct measurements~\cite{Balibar05}.

\begin{figure}[tttt]
\centerline{
  \begin{minipage}[c]{0.55\columnwidth}
\includegraphics[width=\columnwidth]{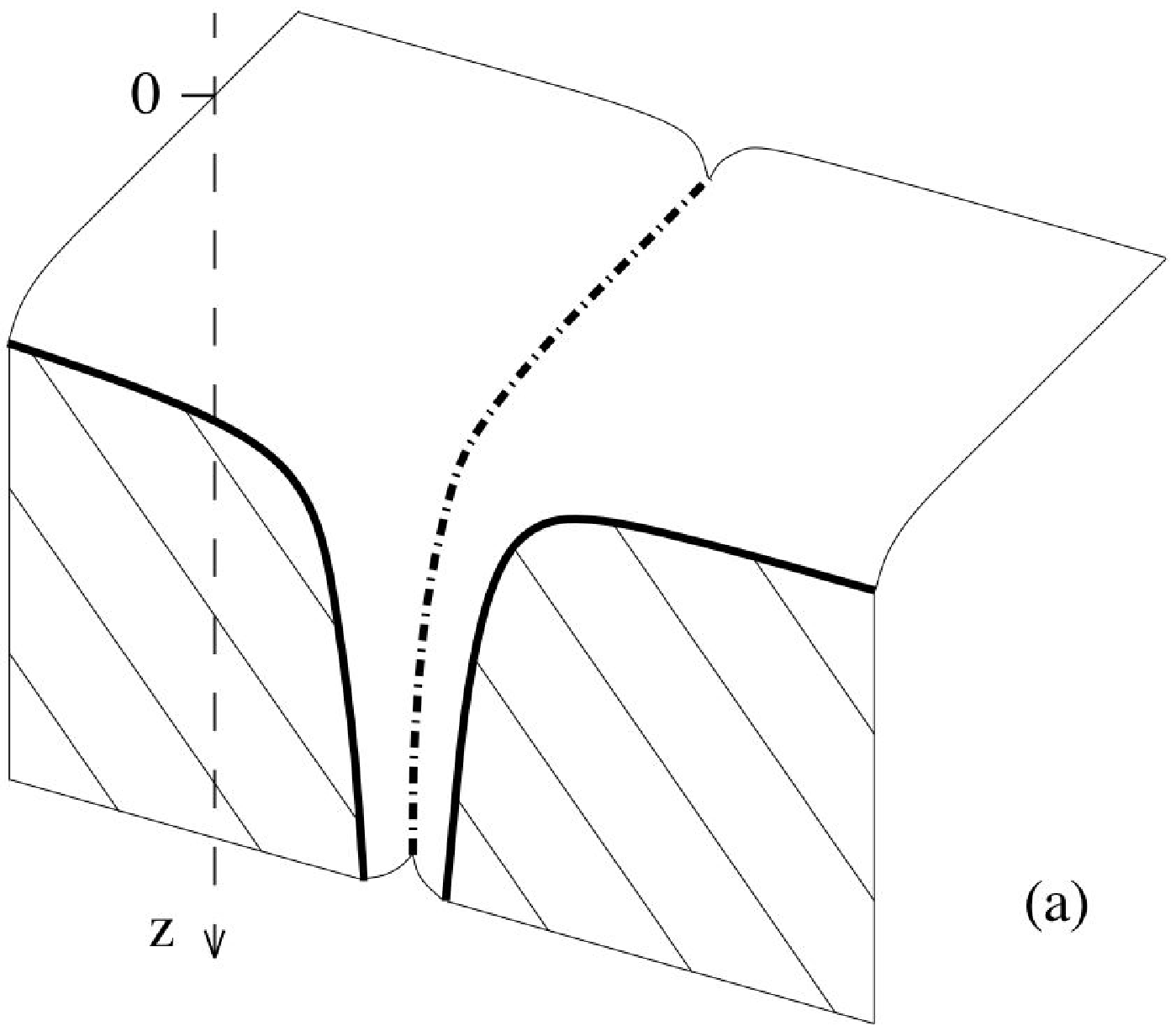}
  \end{minipage}
  \begin{minipage}[c]{0.35\columnwidth}
\includegraphics[width=\columnwidth]{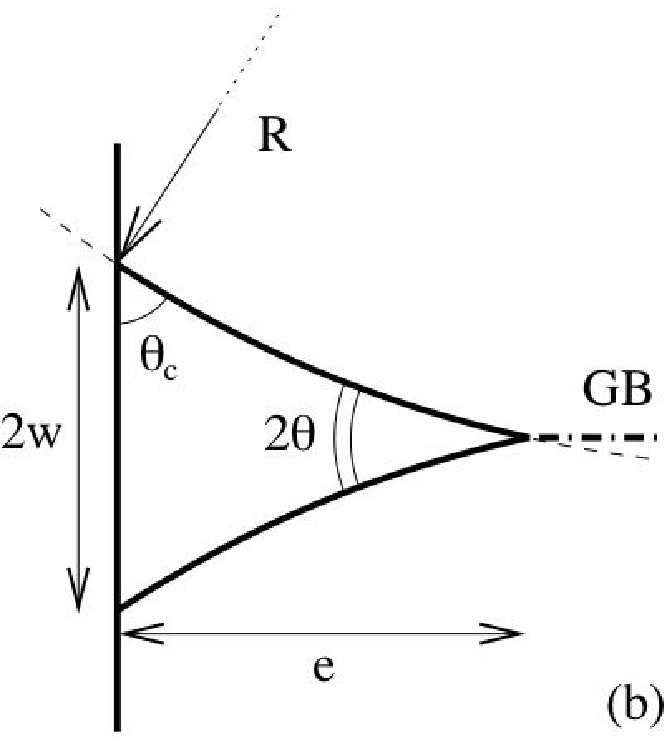}
  \end{minipage}
}
\caption{(a) Three dimensional view of the contact between a grain
boundary (dash-dotted line) and a wall.  The hatched area shows the
contact of the wall with the solid.  (b) Horizontal cross section of
the liquid channel near the wall.}
\label{fig:sketch}
\end{figure}

Figs.~\ref{f1}(a,g) further show that the contact line of a GB with a
wall (here the glass windows) can be wet by the liquid, so that it is
in fact a liquid channel.  We now present a model to calculate the
shape and size of this channel, along with the condition for its
existence.  For simplicity, let us neglect elasticity and assume again
that $\sigma_\mathrm{LS}$ is isotropic.  The force balance on the
contact line between the liquid and the GB then requires this GB to be
a vertical plane perpendicular to the wall, and the liquid channel to
be symmetric with respect to this plane (see Fig.~\ref{fig:sketch}).
Moreover, the LS interface has a contact angle $\theta_\mathrm{c}$
with the glass wall.  We have $\cos \tc =
(\sigma_\mathrm{SW}-\sigma_\mathrm{LW})/\sLS$, where
$\sigma_\mathrm{SW}$ and $\sigma_\mathrm{LW}$ are the solid-wall and
liquid-wall surface tensions, respectively.  In the past,
$\theta_\mathrm{c}$ has been estimated around
45$^\circ$~\cite{Balibar05}.  The present experiment allowed
systematic measurements, using the side copper walls and the glass
plate at the top of the cell.  We found that 45$\textsuperscript{o}$
is indeed an average value, but there is a large scatter in our new
measurements, due to hysteresis and pinning effects~\cite{SasakiJLTP}.

Let $z$ be the depth measured from the top of the grains (Fig.~3(a)) where the
curvature of the LS interface is small so that the pressure $P$ equals
the bulk equilibrium pressure $\Peq$.  Hydrostatic
equilibrium in the liquid gives its pressure 
$P_\mathrm{L}(z) = \Peq + \rho_\mathrm{L} g z$.  If one keeps
neglecting elasticity, the equality of chemical potentials at the LS
interface implies that the pressure in the solid is $P_\mathrm{S}(z) =
\Peq + \rho_\mathrm{S} g z$.  The pressure difference through the LS
interface sets its curvature, according to Laplace's equation.  At a
depth $z$ larger than the capillary length $\lc = \sqrt{\sLS/[ (\rS
-\rL) \, g]}\, \approx $ 1~mm, the curvature in the horizontal plane
dominates, and the LS interfaces in a horizontal plane are circular
arcs of radius $R = \sLS/[\PS (z) - \PL (z)]= {\lc}^2/z$.
Fig.~\ref{fig:sketch}~(b) shows a horizontal cross section of the
channel, and defines its thickness $e$ and width $2w$.  Trigonometry
gives:
\begin{eqnarray}
e(z) = \frac{{\lc}^2}{z} \left( \cos \tc - \sin \t \right) \\ 
w(z) = \frac{{\lc}^2}{z} \left( \cos \t - \sin \tc \right)
\label{eq:ew}
\end{eqnarray}

The liquid channel exists if and only if $\t + \tc < \pi/2$.  This
is why a channel is seen in Fig.~\ref{f1}(a), not in
Figs.~\ref{f1}(c,e).  We checked that the relative correction
due to the vertical curvature is less than 10~\% for $z/\lc>1.7$~\cite{SasakiJLTP}.

\begin{figure}
\centerline{\includegraphics[width=0.95\linewidth]{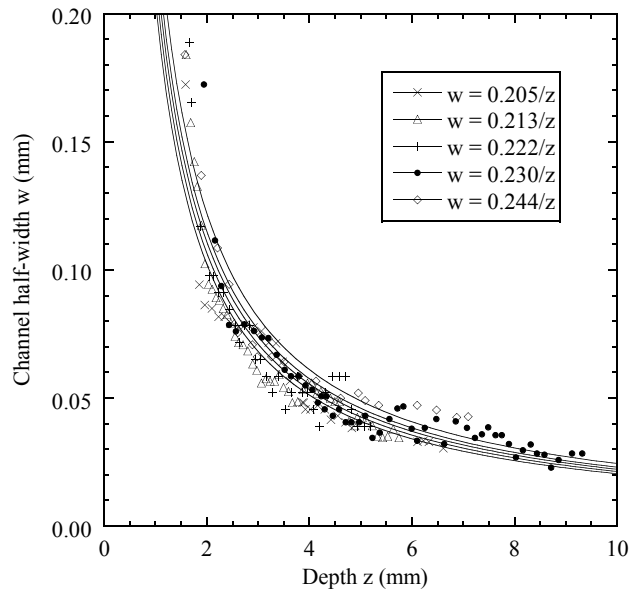}}
\caption{The half-width $w$ as a function of depth $z$ for 5 different 
samples. Good agreement with Eq.~4 is found.}
\label{f5}
\end{figure}

We measured $w$ as a function of $z$ by enlarging
pictures such as Fig.~\ref{f1}(a) and by taking the maximum width of
the gap between homogeneous grey regions on both sides of the channel.
Fig.~\ref{f5} shows fits with $w=A/z$ of 5
different data sets.  The values of $A$ fall in the range from
0.20 to 0.24~mm$^2$ which is consistent with Eq.~(\ref{eq:ew}).  The
channel width vanishes rapidly as $P$ increases: $2w$ reaches
$1\,\mathrm{nm}$ (the typical thickness of a GB~\cite{Pollet}) at
$0.94\,\mathrm{MPa}$ above $\Peq$~\cite{SasakiJLTP}.  Note that we
have neglected elasticity; in a solid with large stress
gradients, there would be positive elastic terms in the solid free
energy, so that liquid channels may exist at $\PL$ much larger
than $\Peq$.  We have also observed that the channels can be pinned to
wall defects which locally favor the presence of the liquid phase.

The present results lead us to reconsider the interpretation of the
experiments by Franck~\etal~\cite{Franck83,Franck85}.  They observed
an apparent wetting of GBs for fcc crystal foams adsorbed on a glass
wall. In our opinion, the above described liquid channels play a role
in the structure of their thin films.  We have shown that the channel
width depends on the departure from $\Peq$; this would explain why the
vertices of the foam widen upon heating, until the grains detach. It
also explains why the contrast is lost when the temperature is lowered
$50-100\,\mathrm{mK}$ below $\Tm$~\cite{Franck85}: it is not due to
the disappearance of the GBs, because the contrast is recovered upon
heating; we rather think that the channels shrink to submicron
dimensions upon cooling, and widen again upon heating.

Let us now reconsider the experiment~\cite{Sasaki06} where we
observed mass transport through solid $^4$He only in the presence
of GBs.  Assuming this transport to take place along the GBs, we had
found that these GBs were superfluid with a critical velocity of order
$1\,\mathrm{m\,s^{-1}}$, a reasonable value for a liquid region a few
atomic layers thick~\cite{Telschow}.  However, we now realize that
mass could also flow along the contact between GBs and walls.  Assuming that
the cross section area of the channel is 
$870\,\mathrm{\mu m^2}$ at a depth $z = 10\,\mathrm{mm}$ below the
free LS interface~\cite{SasakiJLTP}, we find a critical velocity about equal to  
$3\,\mathrm{mm\,s^{-1}}$ along the channel.  This is again a
reasonable value for this channel size~\cite{Wilks}. 
If mass was really
transported along these channels, it would explain why relaxation took
place at least up to 1.1~K while GBs are predicted to
become superfluid only around 0.5~K~\cite{Pollet}.  In order to decide
whether mass flows along the GBs themselves or along the side
channels, we plan measurements in different cell geometries.

Let us finally comment on supersolidity.  Clark et al.  recently
observed anomalies in a TO filled with solid samples grown at constant
temperature from the superfluid liquid~\cite{Clark}, which are likely
to be single crystals, consequently without GBs.  These anomalies may
be due to dislocations whose mobility changes with temperature, as
proposed by de Gennes~\cite{PGG} and suggested by recent
experiments~\cite{Day07}.  Another explanation could be the
superfluidity of dislocation cores~\cite{Boninsegni} but, in this
case, since we have observed no mass transport in single
crystals~\cite{Sasaki06}, we would need to suppose that our crystals,
being at the LS equilibrium, contain a lower dislocation density.

We have also studied samples grown at constant volume in an hour or
so.  As in most TO experiments, we used the ``blocked capillary
method'' to prepare such samples at a final pressure between 26 and 30
bar.  These crystals were always transparent, showing no measurable
light scattering contrary to crystals grown rapidly.  When
depressurized down to the LS equiliibrium, they appeared
polycrystalline with grain sizes of a few microns.  Either they were
already polycrystalline after growth or they were glassy and
crystallized while approaching the melting line.  The possible
supersolidity of such samples could be due to GBs but, in order to
build up a 1\% superfluid density with the inside of superfluid GBs
\textit{only}, a very large GB density would be required.

Could $^4$He samples contain liquid or glassy regions?  It was 
noticed that, after annealing samples grown
rapidly, the cell pressure drops by several
bars~\cite{Grigoriev,Rittner07}.  This means that, before annealing,
these samples contain low density regions which could be liquid
or glassy. We observed a 
definite scattering of light by crystals grown in a few seconds from 
the normal liquid or in 0.1~s from the superfluid, which 
indicates the existence of low density regions with a size comparable 
to the wavelength of visible light. Similar observations were  
made before~\cite{Maekawa,Ford}. 
In their quenched samples,
Rittner and Reppy found very large effects which they presented as
supersolid fractions up to 20~\% or more~\cite{Rittner07}.  In this
case, the large superfluid density could be due to a large
concentration of small liquid regions.  If these regions are connected
by superfluid GBs, the solid samples would show macroscopic phase coherence below
the transition temperature of GBs and below their critical
velocity which might be small.  In any case it is interesting to
proceed with the study of GBs by measuring their critical temperature
and the critical velocity along them.  This should be possible with
experimental cells similar to the one described in this Letter.

We are grateful to N. Jamal, H.J. Maris,  P. Nozi\`eres, and E. Rolley
for helpful discussions.  This work is supported by ANR grant
05-BLAN-0084-01.


\begin{thebibliography}{0}
\bibitem{note}In reality, the best fit gave slightly different values 
of $\sLS$ and $\theta$ on each side of the GB. Eq.~(1) should thus 
be written as $\sGB = 
\sLS^1\cos{\theta_{1}} + \sLS^2\cos{\theta_{2}}$, with 1 and 2 
standing for each side. We have taken 
$\theta = (\theta_{1} + \theta_{2})/2$.
\end{thebibliography}
\end{document}